\documentclass{article}

\usepackage{arxiv}

\usepackage[utf8]{inputenc} 
\usepackage[T1]{fontenc}    
\usepackage{hyperref}       
\usepackage{url}            
\usepackage{booktabs}       
\usepackage{amsfonts}       
\usepackage{nicefrac}       
\usepackage{microtype}      
\usepackage{lipsum}		
\usepackage{graphicx}
\usepackage{natbib}
\usepackage{doi}
\usepackage{multirow}

\title{2D MULTI-CLASS MODEL FOR GRAY AND WHITE MATTER SEGMENTATION\\
OF THE CERVICAL SPINAL CORD AT 7T}

\date{October 12, 2021}	

\author{ \href{https://orcid.org/0000-0003-0677-8991}{\includegraphics[scale=0.06]{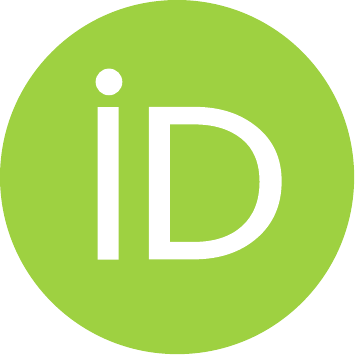} \hspace{1mm}Nilser J. Laines Medina}\\
	Aix Marseille Univ, CNRS, CRMBM, Marseille, France\\
	APHM, CHU Timone, Pôle d’Imagerie Médicale, CEMEREM, Marseille, France \\
	\texttt{nilser-jherald.LAINES-MEDINA@univ-amu.fr} \\
	\And
	
	\href{https://orcid.org/0000-0003-4318-0024}{\includegraphics[scale=0.06]{orcid.pdf}\hspace{1mm}Charley Gros} \\
	NeuroPoly Lab, Institute of Biomedical Engineering, Polytechnique Montréal, Montréal, QC, Canada \\
	MILA - Québec AI Institute, Montréal, QC, Canada \\
	\texttt{charley.gros@gmail.com} \\
	
	\And
	
	\href{https://orcid.org/0000-0003-3662-9532}{\includegraphics[scale=0.06]{orcid.pdf}\hspace{1mm}Julien Cohen-Adad} \\
	NeuroPoly Lab, Institute of Biomedical Engineering, Polytechnique Montréal, Montréal, QC, Canada \\
	MILA - Québec AI Institute, Montréal, QC, Canada \\
	\texttt{jcohen@polymtl.ca} \\
	
	\And
	
	\href{https://orcid.org/0000-0003-0850-1742}{\includegraphics[scale=0.06]{orcid.pdf}\hspace{1mm}Virginie Callot} \\
	Aix Marseille Univ, CNRS, CRMBM, Marseille, France\\
	APHM, CHU Timone, Pôle d’Imagerie Médicale, CEMEREM, Marseille, France \\
	\texttt{virginie.callot@univ-amu.fr} \\
	
	\And
	
	\href{https://orcid.org/0000-0002-7897-9672}{\includegraphics[scale=0.06]{orcid.pdf}\hspace{1mm}Arnaud Le Troter} \\
	Aix Marseille Univ, CNRS, CRMBM, Marseille, France\\
	APHM, CHU Timone, Pôle d’Imagerie Médicale, CEMEREM, Marseille, France \\
	\texttt{arnaud.le-troter@univ-amu.fr} \\
}

\hypersetup{
pdftitle={A template for the arxiv style},
pdfsubject={q-bio.NC, q-bio.QM},
}

\begin{document}
\maketitle

\begin{abstract}
The spinal cord (SC), which conveys information between the brain and the peripheral nervous system, plays a key role in various neurological disorders such as multiple sclerosis (MS) and amyotrophic lateral sclerosis (ALS), in which both gray matter (GM) and white matter (WM) may be impaired. 
While automated methods for WM/GM segmentation are now largely available, these techniques, developed for conventional systems (3T or lower) do not necessarily perform well on 7T MRI data, which feature finer details, contrasts, but also different artifacts or signal dropout.

The primary goal of this study is thus to propose a new deep learning model that allows robust SC/GM multi-class segmentation based on ultra-high resolution 7T T2*-w MR images. The second objective is to highlight the relevance of implementing a specific data augmentation (DA) strategy, in particular to generate a generic model that could be used for multi-center studies at 7T.
	
\end{abstract}


\section{INTRODUCTION}
\label{sec:intro}

MRI of the human spinal cord (SC) is a routine clinical procedure that is conventionally performed at a magnetic field strength of 1.5 or 3 Tesla. Recently, Ultra-High Field (UHF) scanners (at 7T \cite{barry_spinal_2018} or higher \cite{geldschlager_ultrahigh-resolution_2021}) have shown great potentialities over conventional systems, as they offer better signal-to-noise ratio (SNR) and contrast-to-noise ratio (CNR) that could be traded for better spatial resolution and increased detection of anatomical and pathological features. Consequently, 7T SC MRI has become an interesting tool for research aimed at improving both clinical diagnosis and description of white matter and gray matter (WM/GM) cord abnormalities. 
To extract WM/GM information, the segmentation process is an essential method. The two boundaries to be delineated are the WM/CSF (cerebrospinal fluid, surrounding to the cord) and WM/GM (external contour of GM) interfaces. As manual segmentation is time consuming, especially for GM, and has been demonstrated on a very large multi-center study not to be very reproducible, both inter- and intra-experts (Prados et al. \cite{prados_spinal_2017}), several fully automated segmentation techniques have been proposed to the community over the last ten years. In their literature review of 2016, De Leener et al \cite{de_leener_segmentation_2016} detailed the methods available at that time, most of which were based on atlases and registration \cite{de_leener_pam50_2018}\cite{massire_anterior_2020} or label fusion \cite{asman_robust_2013}. More recently, and with the emergence and success of deep-learning (DL) approaches that are increasingly used for medical image analysis, and especially for MRI segmentation tasks, several models derived from U-Net \cite{ronneberger_u-net_2015} architectures have proven to be very efficient and robust for GM \cite{perone_spinal_2018} and SC segmentation \cite{gros_automatic_2019}.

Most of these methods and models have been made available open-source within the spinal cord toolbox (SCT) \cite{de_leener_sct_2017}, and hence allow the community to contribute to new models that can be implemented quickly on an optimized architecture for SC. It is worth noting that SCT is additionally compatible with the PyTorch models implemented in the recent ivadomed toolbox \cite{gros_ivadomed_2021}, an open-source python package used for the design, end-to-end training and evaluation of new DL models applied to medical imaging data. One particularly useful feature in ivadomed is the data loader that can parse participants’ image and metadata for custom data splitting or additional information during training and evaluation, provided that datasets are organized according to the Brain Imaging Data Structure (BIDS) \cite{gorgolewski_brain_2016}.

Nonetheless, most of the segmentation methods proposed so far and mentioned above have been widely applied and validated in the context of 3T MRI. With the emergence of 7T systems, SC images with ultra-high spatial resolution (UHR) and contrasts are collected, revealing new details, such as nerve roots or blood vessels, and artifacts or signal drop out, that may challenge existing segmentation methods.
The main objective of this study was thus to propose a dedicated approach for 7T SC MRI segmentation, based on an optimized ivadomed model to be integrated in SCT (via the new \textit{sct\_deepseg} function).

The proposed model is based on T2*-weighted images with different spatial resolutions from healthy and pathological subjects. The resulting variability introduced in the MRI database, cumulated with different variants of data augmentation, aimed at avoiding overfitting during the training. A feasibility study was conducted to highlight the relevance of implementing a specific data augmentation strategy that would be robust in the presence of artifacts or intrinsic variability of multi-center acquisitions.

\section{MATERIALS AND METHODS}
\label{sec:pagestyle}

\subsection{Participants}
\label{ssec:subhead}
A total of 77 subjects were included in this retrospective study. The imaging protocol was approved by local Ethics Committees and written consents were obtained from all subjects prior to MR examinations.
\subsection{Monocentric and multicentric datasets}
\label{ssec:subhead}
A large monocentric dataset (DS1) was first constructed for development purpose using 72 subjects, including 34 Healthy Controls (HC), 25 patients with Amyotrophic Lateral Sclerosis (ALS), 13 patients with Multiple Sclerosis (MS), scanned in Marseille (France) using a 7T Magnetom system (Siemens, Erlangen, Germany) and an 8-channel Tx/Rx neck coil (Rapid Biomedical, Rimpar, Germany). A second smaller multicentric dataset (DS2), used to assess the model ability to generalize to new acquisition conditions, was composed of 5 HC, acquired in 3 different 7T MR scanners/centers (Zurich (7T Terra, Rapid Biomedical coil), New York (7T Magnetom, home-made volume coil), and Marseille (7T Terra, Rapid Biomedical coil)).
\subsection{Image acquisition}
\label{ssec:subhead}
All images were acquired using a 2D multi-echo T2*-weighted GRE sequence. Slices were positioned from C1 to C7, in the axial plane, perpendicular to the cord. High, Medium and Low-resolution series (HR, MR, LR, respectively) were acquired (depending on time constraint and investigation purpose). Acquisition parameters for HR are detailed in \cite{massire_high-resolution_2016}. Table ~\ref{DS1DS2tab} describes the total number of slices and subjects acquired for each resolution, subject category, and datasets.

\begin{table}
\centering
\begin{tabular}{|c|c|c|c|c|c|c|} 
\hline
DS & Acq & Res & Th & Matrix & N & Group\\
\hline
\multirow{4}{*}{DS1} & HR & $0.17^2$ & 2.2 & 832x1024 & 26 & (78,0,89) \\
\cline{2-7}
& HR & $0.20^2$ & 4 & 832x1024 & 1 & (13,0,0) \\
\cline{2-7}
& MR & $0.27^2$ & 5 & 468x576 & 42	& (267,283,0) \\
\cline{2-7}
& LR & $0.40^2$ & 3.3 & 348x348 & 3 & (15,0,0) \\
\hline
\multirow{4}{*}{DS2} & MR & $0.26^2$ & 6 & 500x500 & 1 & (12,0,0) \\
\cline{2-7}
& MR & $0.30^2$ & 6.6 & 304x512 & 1 & (9,0,0) \\
\cline{2-7}
& LR & $0.40^2$ & 2.2 & 384x384 & 1	& (7,0,0) \\
\cline{2-7}
& MR & $0.27^2$ & 2.5 & 468x576 & 2 & (34,0,0) \\
\hline
\end{tabular}
\caption{\label{DS1DS2tab}Dataset (DS) distribution by acquisition type (Acq HR,MR,LR), in-plane resolution (Res in mm2), slice thickness (Th in mm), Matrix size, number of subjects (N), number of 2D slices per Group (HC,ALS,MS).}
\end{table}

To increase signal- and contrast-to-noise of anatomical images, Multi Echo Data Image Combination (MEDIC) images were generated from multi-echo T2*-w series by computing the sum of squares of the image from each echo. The mean$\pm$stdev number of slices per subject was 13.9$\pm$5.9. SC and GM masks were manually segmented for each slice when possible, by two different raters (MS patient group the first rater (XX), and ALS patient group for the second (XX), with an equivalent split for the HC group).

\subsection{Pre-processing for bids dataset organization}
\label{ssec:subhead}

\begin{figure}[htb]
	\centering
	\includegraphics[]{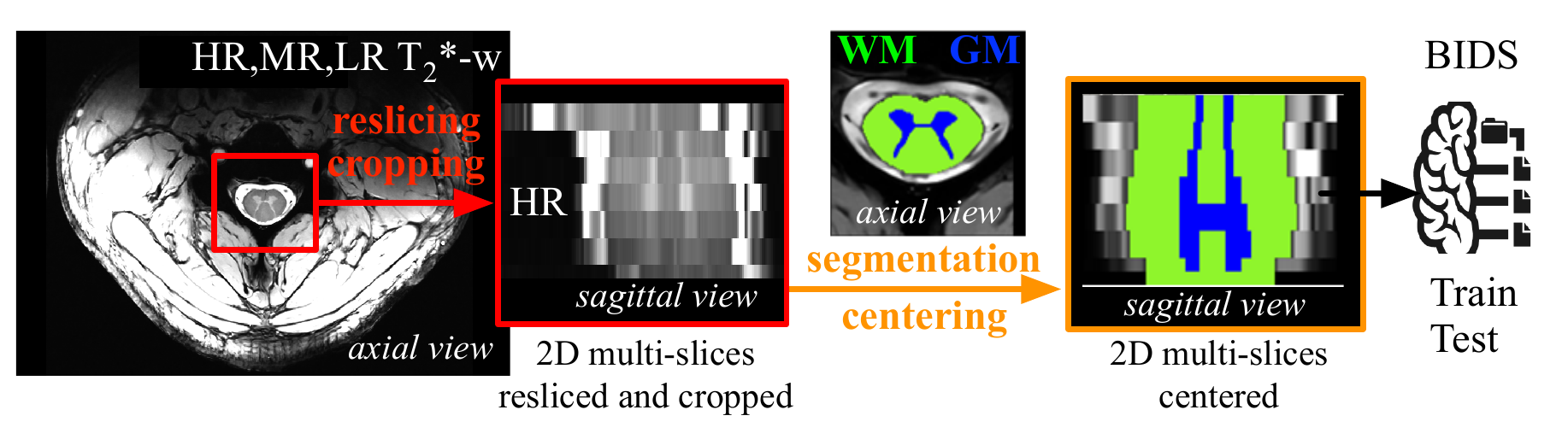}
	\caption{Pre-processing pipeline for data organization}
	\label{fig1}
\end{figure}

A pipeline was developed for data BIDS organization (Fig.~\ref{fig1}). Each T2*-w slice was resliced at the highest resolution HR of $(0.175 mm)^2$ and cropped using a bounding box of $128\times128$ pixels $(22.5 mm)^2$ centered in the SC area. In this present study, the centering step was done using the barycenter of the mask of manual segmentation but will have to be automated in the future. A linear interpolation was applied during the reslicing step.

Finally, the complete database was organized in the BIDS convention, with one folder per subject containing the stacked anatomical T2*-w images and the corresponding segmentation (Ground Truth$=$GT) with binary masks of the spinal cord (SC$=$WM$+$GM) and GM.

A total of 807 slices (out of the 1056 anatomical images initially acquired) were processed and included in the BIDS database, separated into the 2 datasets (745 images for DS1 and 62 images for DS2 respectively). The remaining (249 images), corresponds to data that could not be segmented by the experts because of important artifacts or a too lower GM/WM contrast that would induce an uncertainty in the depiction of tissue boundaries. On these criteria, a total of 6.4\% for SC and 23.4\% for GM were rejected.

\subsection{Post-processing and experimental designs}
\label{ssec:subhead}

In a first experimental design (Exp1), the monocentric dataset DS1 was used to assess the relevance of simultaneously optimizing a single model with multi-class segmentation (MCS), as compared to two separate single-class segmentation models (SCS), for the automatic segmentation of SC and GM classes.
In a second experimental design (Exp2), different Data Augmentation (DA) strategies were compared using DS1 and the multicentric dataset DS2. Two pre-trained segmentation models for GM and SC regions from SCT, respectively run by the functions \textit{sct\_deepseg\_gm} \cite{perone_spinal_2018} and \textit{sct\_deepseg\_sc} \cite{gros_automatic_2019}, were additionally used to perform an automatic segmentation, considered as a reference model, called “SCT” in this study.

\subsubsection{Exp1: Single-class (x2) vs Multi-class (x1) models}
\label{sssec:subsubhead} 

\begin{figure}[htb]
\centering
\includegraphics[height=5cm]{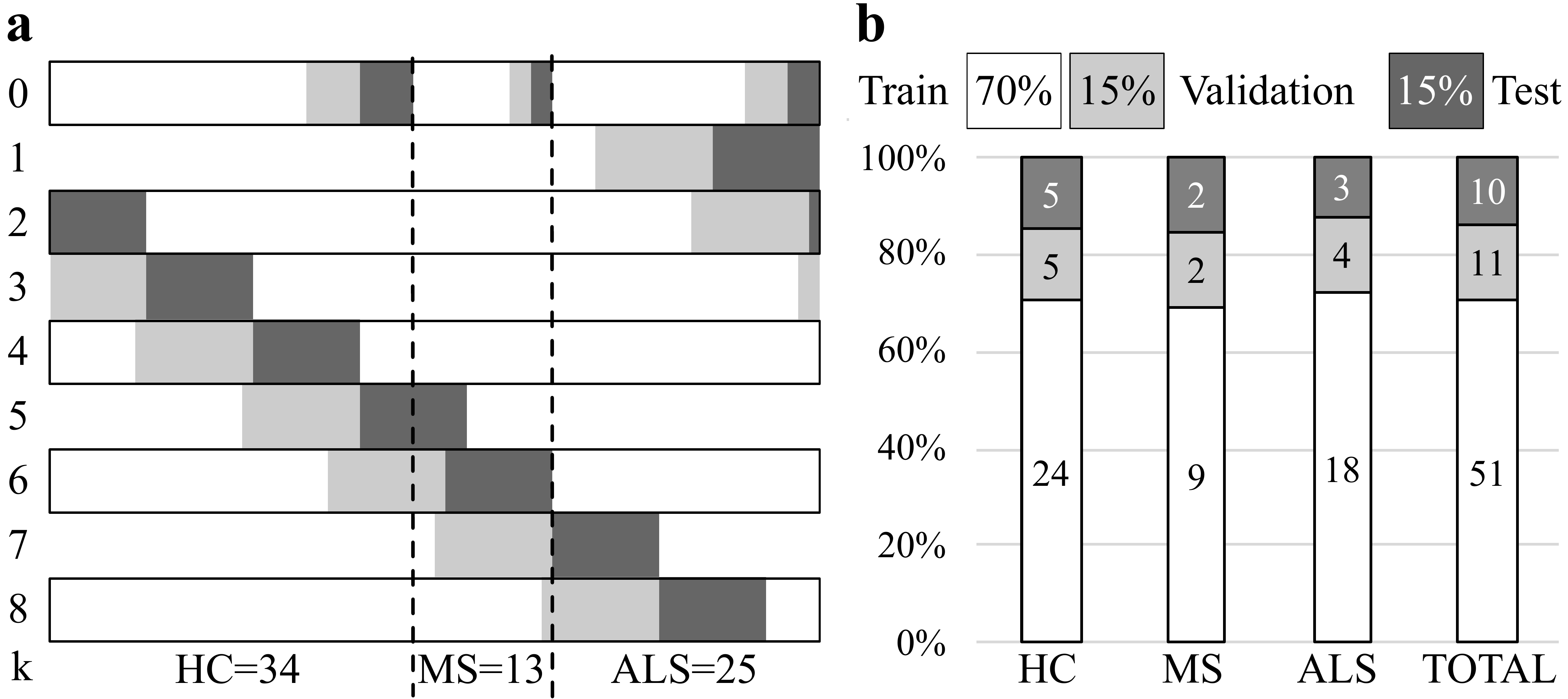}
\caption{a. k-folding configuration for Cross-Validation (k=9), b. Subject numbers per group (HC,MS,ALS) for the fold 0}
\label{fig2}
\end{figure}

To compare the single-class and multi-class models, a k-fold validation method was adopted Fig.\ref{fig2}. This approach consists in a Cross-Validation (CV) method that has been demonstrated to be robust and efficient when the sample is limited. For this, 9 folds were established. For each fold (from 0 to 8), the data were distributed as recommended in the literature, i.e. a split corresponding approximately to 70\% for the Train set, 15\% for the Validation set and 15\% for the Test set. Fig.\ref{fig2}a gives the schematic representation of distribution of the 9 folds.

An approximation in the constitution of the sets (not always containing the same number of images) had to be performed, due to the fact that the number of subjects per group and the number of slices per subjects were neither equals nor proportional to the 70/15/15 ratio. The constraint was that for each fold, no slice of the same patient could appear in 2 different sets among Train/Validation/Test. The fold 0 was performed taking into account the heterogeneity of the subject group, with the respect of the 70/15/15 proportion for each category (HC,MS,ALS), as detailed in Fig.\ref{fig2}b.

\subsubsection{Exp2: Data Augmentation (DA) strategies}

\begin{figure}[htb]
\centering
\includegraphics[height=7cm]{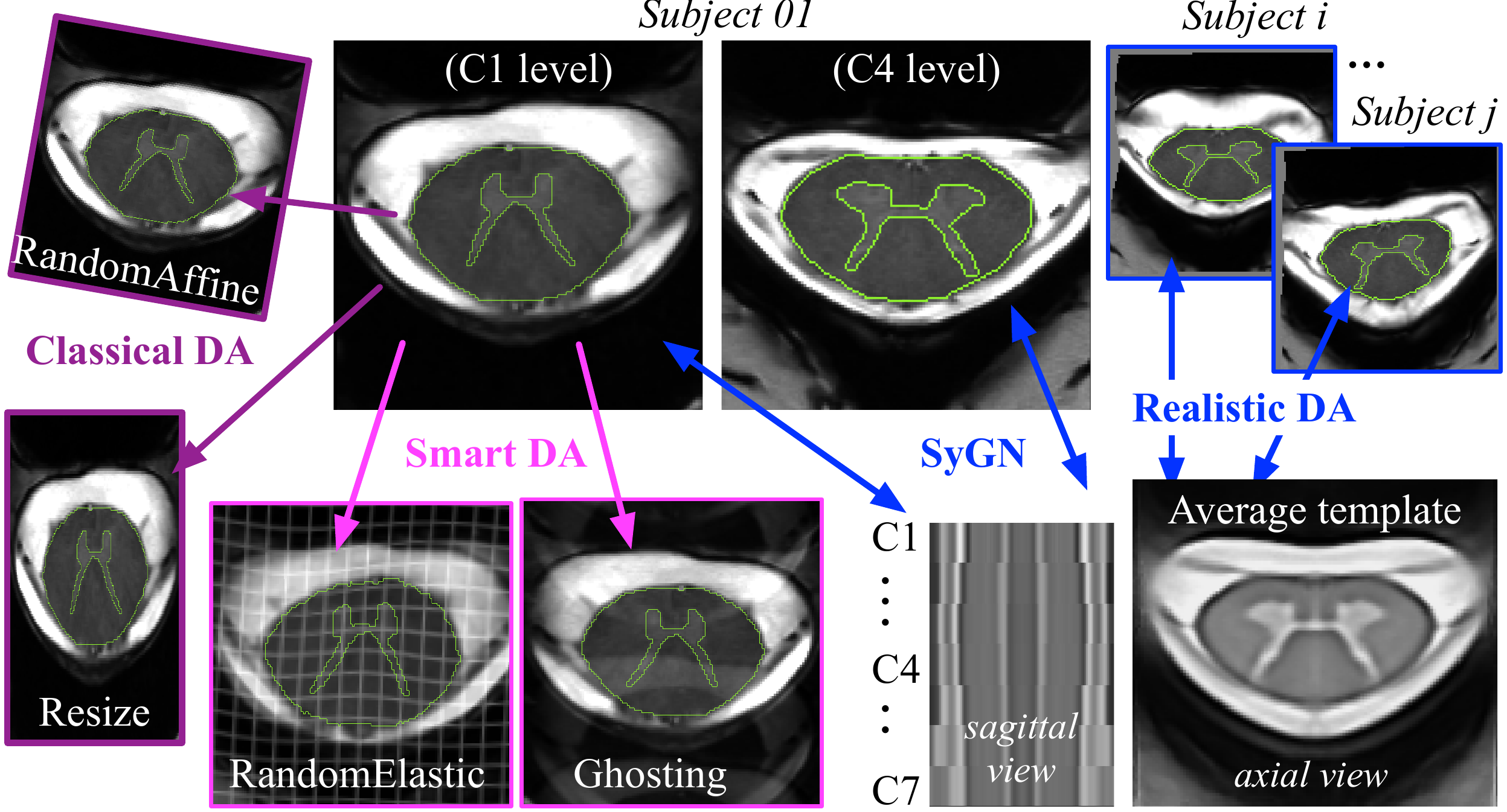}
\caption{Six samples of HYBRID\_DA for one single subject (C1 and C4 cervical levels of the \textit{Subject 01}): a Random Affine and a Resize transformation (Classical DA, purple color), a Random Elastic and a Ghosting effect (Smart DA, pink color), and a realistic augmentation on \textit{Subjects i and j} (Realistic DA, blue color).}
\label{fig3}
\end{figure}

This study also aimed at exploring the impact of data augmentation (DA) for the generalization of the best multi-class model (MCS) (here fold-0, see Results part). 
For this purpose, different DA strategies were evaluated on DS1 and DS2, with an original contribution consisting in a new approach (called "Hybrid\_DA") based on a set of realistic geometric transformations ("Realistic DA"), combined with both "Smart DA" and "Classical DA" strategies (Fig.\ref{fig3}).
The TorchIO python Library \cite{perez-garcia_torchio_2021} was used to generate a set of "Smart" transformations, mimicking MRI-specific effects such as ghosting effect and motion artifacts (Fig.\ref{fig3}, pink color). The classical DA (such as resize or random affine transformations) was obtained using the default implementation in ivadomed (Fig.\ref{fig3}, purple color).
Our “Realistic DA” (Fig.\ref{fig3}, blue color) was constructed by the estimation of inter-subject registrations using Symmetric Group-wise Normalization approach (SyGN, \cite{avants_optimal_2010}\cite{avants_multivariate_2007}), and the use of the script \textit{antsMultivariateTemplateConstruction.sh} from ants Library \cite{avants_insight_2014}. The resulting template space made it possible to compose several inter-subject deformation fields. The process more specifically consisted in morphing a slice of the \textit{Subject i} into a slice of the \textit{Subject j}, thus mixing the contrast of a subject with the anatomy of another subject (and vice versa). For each slice of all N subjects, a set of (N-1) slices with realistic contrasts and correct anatomy were thus generated.

To compare the performance of our Hybrid\_DA strategy, other models were evaluated, including Without\_DA (no augmentation), Classical\_DA (DA implemented by default in ivadomed, like random affine transformations) and SCT (GM and SC models in SCT).

\subsection{Statistics and evaluation metrics }

To evaluate all segmentations compared to manual segmentation (GT), the Dice Similarity Coefficient (DSC), the Hausdorff distance (HDRFDST), and the Volumetric Similarity (VOLSMTY) were computed on each 2D slice, using the “EvaluateSegmentation” Tool \cite{taha_metrics_2015}. DSC values of each method were evaluated through a one-way analysis of variance (Nonparametric Wilcoxon test, with a level of significance defined as $p<0.05$, corrected for multiple comparisons) were used to compare performance among each method. The latter included a quantification of DSC outliers, which was carried out considering the 95\% confidence intervals for the mean and with a 15\% of trimming (for the calculation of truncated mean values). 

\section{RESULTS AND DISCUSSION}
\label{sec:pagestyle}
\subsection{SCS vs MCS using k-fold-CV (Exp1)}
\label{ssec:subhead}
\begin{figure}[htb]
\centering
\includegraphics[height=9cm]{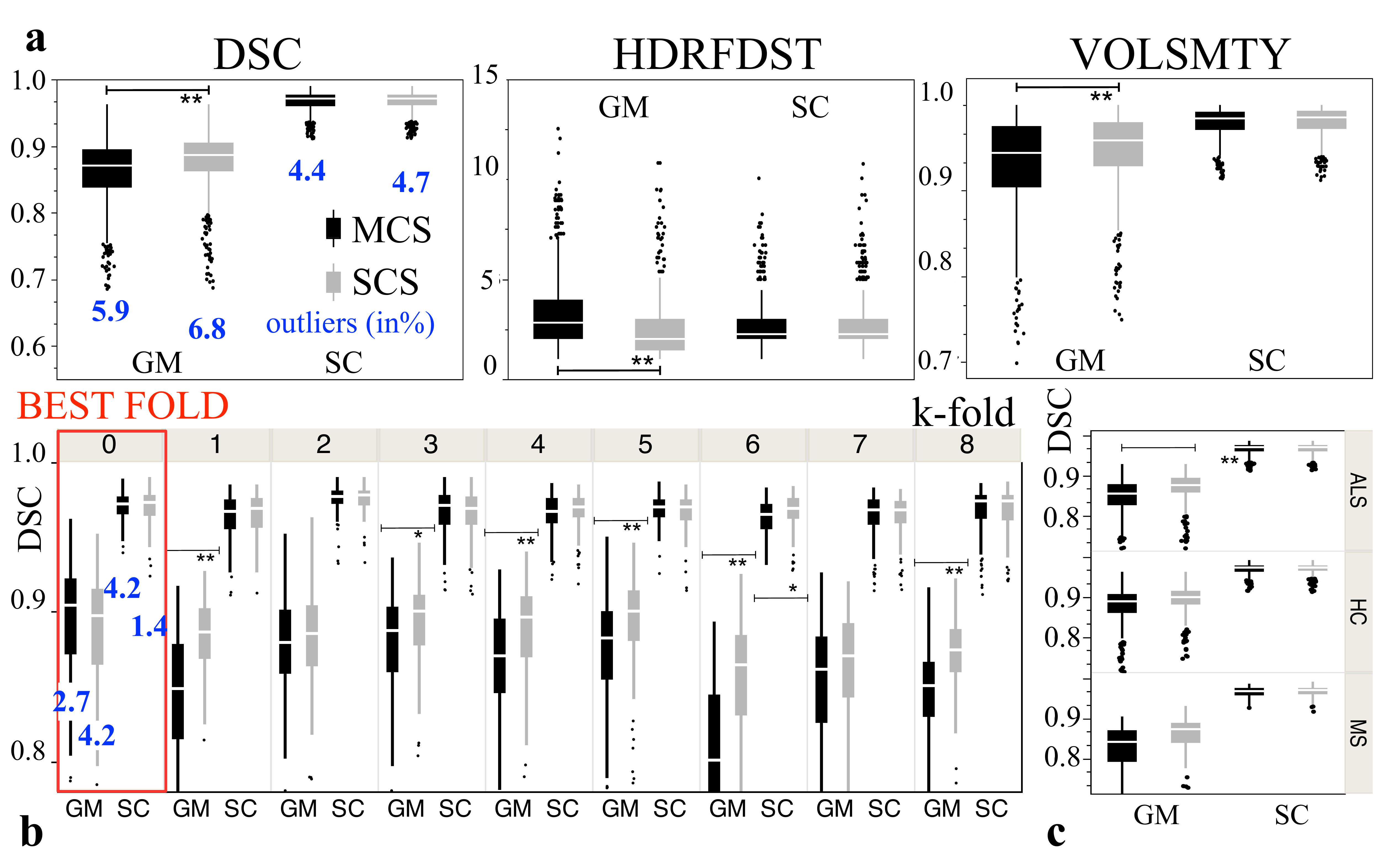}
\caption{a. Value distributions of all metrics (Dice Similarity Coefficient (DSC), Hausdorff distance (HDRFDST in pixel), and Volumetric Similarity (VOLSMTY)) per class (GM and SC) for all folds 
for both Multi-Class (MCS, in black) and Single-Class Segmentation (SC, in gray) approaches. b. DSC per class (GM/SC) for each fold. The best fold is highlighted in red (fold 0). c. Mean DSC for each group (ALS,HC,MS). 
}
\label{fig4}
\end{figure}

Using all testing sets of DS1 and considering all folds, significantly higher DSC, VOLSMTY and lower HDRFDS values (corrected for multiple comparisons ** $p<0.0001$) were observed in average in SCS compared to MCS for GM class, without significant difference for SC class (see Fig.\ref{fig4}a). Truncated mean DSC values were $0.88{\pm}0.04$ vs $0.86{\pm}0.05$ for GM, and $0.97{\pm}0.01$ vs $0.97{\pm}0.01$ for SC, for SCS and MCS, respectively (all outliers excluded). Seven out of 9 folds showed the same result (**SCS$>$MCS) (see Fig.\ref{fig4}b) and similar observation was made for each subject category (ALS, HC, MS) (see Fig.\ref{fig4}c). However, the best fold (estimated on the fold 0, highlighted in red on Fig.\ref{fig4}b) presented with no difference between SCS vs MCS ($0.89{\pm}0.05$ and $0.89{\pm}0.04$ for GM and $0.97{\pm}0.01$ and $0.97{\pm}0.01$ for SC), respectively, with slightly higher results for MCS. For all folds, the percentages of outliers were 6.8\% and 5.9\% for GM and 4.7\% and 4.4\% for SC, respectively for SCS and MCS.

\subsection{Impact of DA and comparison with existing SCT models (Exp2)}
\begin{figure}[htb]
\centering
\includegraphics[height=8cm]{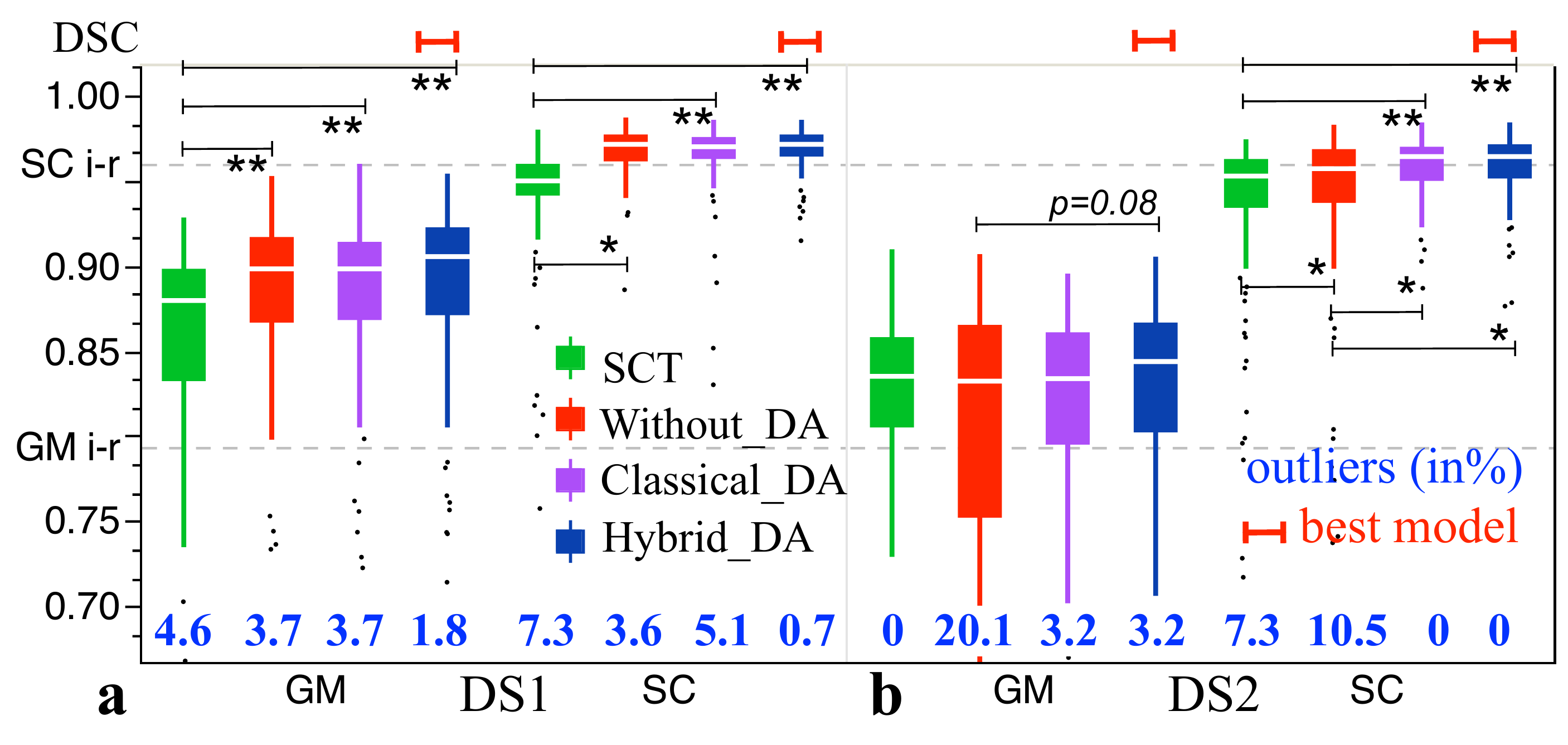}
\caption{Boxplot representations of Dice Similarity Coefficient (DSC) values obtained with the different DA approaches (Without, Classical, Hybrid) and the reference SCT models, for both GM and SC classes. Testing was computed only on previously defined fold 0 for DS1 (on left), as well as on DS2 (on right). DSC inter-raters values (i-r) are reported.}
\label{fig5}
\end{figure}

For fold 0 on DS1, significantly higher DSC values were observed (cf. Fig.\ref{fig5}a) for all proposed DA approaches compared to SCT models, for both GM and SC classes. Surprisingly, the Without\_DA model showed high performance on DS1, but as expected from the state-of-the art, collapsed on DS2 (lower mean and higher standard deviation and outliers (cf. Fig.\ref{fig5}b)), which reinforces the interest of DA to increase robustness and built a model more generalizable.

For DS1, truncated mean DSC values for SCT, Without\_DA, Classical\_DA and Hybrid\_DA models were respectively $0.85{\pm}0.07$, $0.87{\pm}0.13$,$0.87{\pm}0.13$ and $0.88{\pm}0.08$ for GM and $0.94{\pm}0.04$, $0.95{\pm}0.12$, $0.95{\pm}0.13$ and $0.97{\pm}0.04$ for SC, and for DS2 $0.82{\pm}0.05$, $0.70{\pm}0.27$,$0.87{\pm}0.13$ and $0.88{\pm}0.08$ for GM and $0.94{\pm}0.04$, $0.95{\pm}0.12$, $0.95{\pm}0.13$ and $0.97{\pm}0.04$ for SC.

\section{CONCLUSION}
\label{sec:pagestyle}
This study highlights the interest of training a specific model containing images of healthy controls and patients at different spatial resolutions for the segmentation of UHR images acquired at 7T, showing very high performances (the best model Hybrid\_DA trained on the best fold gives a truncated mean DSC=$0.88{\pm}0.08$ for GM, and DSC=$0.97{\pm}0.04$ for SC), significantly better than the existing models trained on images acquired at 1.5T and 3T.

Single-class models showed in average a better performance than multi-class models, especially for GM segmentation, but required two trainings instead of one and especially two inferences. Nonetheless, the best performances (higher similarity metrics, faster training time and lower percentage of outliers) were ultimately obtained using the multi-class segmentation (fold 0). This model was enriched by integrating a hybrid data augmentation (composed of "classical" geometric transformations, artifacts and real GM/WM contrasts distorted with anatomically constrained deformation fields), inspired by the recent study of Zhao et al.\cite{zhao_data_2019} which addresses the problem of realistic aspect of augmented data using a transformation learned by a convolutional neural network. Further testing and applications on external data are now required to fully evaluate the added value of our approach. In that perspective, our best model was added as the Spinal Cord Toolbox (SCT) and made publicly available on the \href{https://github.com/ivadomed/model\_seg\_gm-wm\_t2star\_7t\_unet3d-multiclass}{ivadomed github repository}.

\section{Acknowledgments}
\label{sec:acknowledgments}
None of the authors have potential conflicts of interest to disclose.
The authors sincerely thank R. Dintrich and S. Demortière from CRMBM-CEMEREM (Marseille, France) for SC/GM manual segmentation, as well as M. Seif and J. Vannesjö from Balgrist University Hospital (Zürich, Switzerland) and A. Seifert from the Biomedical Engineering and Imaging Institute (Mount Sinai, NY, USA) for providing external 7T SC MR dataset. This work was performed in a laboratory member of France Life Imaging network (grant ANR-11-INBS-0006) and supported by the French Centre National de la Recherche Scientifique, the Marseille Imaging Institute, the NeuroMarseille Institute, ARSEP (Fondation pour l’Aide à la Recherche sur la Sclérose En Plaques), Fondation Thierry Latran, and MarMaRa Institute (Marseille Maladies rares).

\section{Compliance with ethical standards}
\label{sec:ethics}

The study was approved by the local ethics committee (Comité de Protection des Personnes Sud-Méditerranée I, ID RCB : 2011-A00929-32), and written informed consent was obtained before MR imaging for all subjects.

\bibliographystyle{unsrtnat}
\bibliography{references} 

\end{document}